\documentclass[aps,superscriptaddress,twocolumn,showpacs,preprintnumbers,amsmath,amssymb]{revtex4}
\usepackage[cp1251]{inputenc}
\usepackage[english]{babel}
\usepackage{amssymb,amsmath}
\usepackage{amstext} 
\usepackage[dvips]{hyperref}
\usepackage[mathscr]{eucal}
\usepackage{longtable}
\usepackage{graphicx}% Include figure files
\begin{document}
\title{QED Model of Resonance Phenomena  in Quasionedimensional Multichain Qubit Systems with Axial Symmetry}
\author{ Alla Dovlatova (a), Dmitry Yearchuck (b)\\
\textit{(a) - M.V.Lomonosov Moscow State University, Moscow, 119899, \\ (b) -  Minsk State Higher Aviation College, Uborevich Str., 77, Minsk, 220096, RB; yearchuck@gmail.com,}}
\date{\today}% It is always \today, today,
\begin{abstract}The analytical solution of the task of the interaction of quantized EM-field with multichain quasionedimensional axially symmetric qubit system by taking into account both the intrachain and interchain qubit coupling has been obtained for the first time. The appearance of additional lines in optical spectra of quasionedimensional systems  strongly interacting with EM-field is predicted. They are result of nonstationary registration conditions, to be consequence of Rabi wave packet formation and correspond to Fourier transformation of revival part of Rabi wave packets in temporal dependence of the integral inversion.  It is argued the applicability of Su-Schrieffer-Heeger-model of organic conductors for perfect quasionedimensional carbon zigzag shaped nanotubes   to be consequence of 2D-1D transition  with diameter decrease.  \end{abstract}
\maketitle 
\section{Introduction}
 Quantum electrodynamics (QED) consideration of interaction of electromagnetic (EM) field with matter is used at present very rare in practical applications, for instance, quantum nature  of EM-field is not taken into account by its interaction with nanoobjects like to carbon nanotubes, both in theoretical and experimental aspects \cite{Dresselhaus}, \cite{Bose}. There exist at present in QED-theory the analytical solutions of the task of the interaction of quantized single mode                                                                            EM-field with one qubit (two level system), Jaynes-Cummings model (JCM), with multiqubit systems without interaction between qubits, Tavis-Cummings model,  and  the model described in \cite{Slepyan_Yerchak}, generalizing  Tavis-Cummings model  by taking into account the 1D-coupling between qubits. 
The model to be considered continues given  series, that is, it is the model of the interaction of quantized EM-field with quasionedimensional axially symmetric multichain qubit system by taking into account both the intrachain and interchain qubit coupling. Having in view the practical applications, the quasi-1D zigzag-shaped carbon nanotubes (ZSCNTs) will represent multichain qubit system in the task aimed. Simultaneously the quasi-1D-model of  ZSCNTs for description of electronic properties will be also proposed, since existing at present models of NTs  are 2D-models and two dimensional lattice structure of a single wall carbon nanotube (SWNT)
is specified uniquely by the chirality defined by two integers
$(n,m)$ \cite{Dresselhaus}, \cite{M.Dresselhaus}, \cite{Saito}.
 In the Raman spectra of a $2D$-SWNT are  two in-plane $G$ point longitudinal
and transverse optical phonon ($LO$ and  $TO$) modes \cite{Reich} and
the out-of-plane radial breathing mode (RBM) \cite{Dresselhaus M.S} are observed. It is shown in \cite{Sasaki}, that RBM-mode has chirality dependent frequency shift  in metallic carbon $2D$-SWNT.
The $LO$ and $TO$ phonon modes at
the $G$ point in the two-dimensional Brillouin zone
are degenerate in graphite and graphene, however they split in $2D$-SWNT into two peaks,
denoted by $G^+$ and $G^-$peaks, respectively, \cite{G.Dresselhaus} because of the
curvature effect. The agreement of experimental Raman studies of NTs with diameter $ > 1 nm$ with $2D$-SWNT-theory unambiguously indicates, that given NTs, produced  by CVD are really $2D$-systems.  At the same time the results reported in  \cite{Erchak}, \cite{Ertchak}, \cite{Ertchak_Stelmakh} show, that ZSCNTs, produced by $\langle 111 \rangle$ high energy ion implantation (HEIM) of diamond single crystals   are  quasi-$1D$ systems. Given conclusion is true for the narrow  tubes with diameter $ < 1 nm$ at all. Really, along with results, reported in  \cite{Erchak}, \cite{Ertchak}, \cite{Ertchak_Stelmakh} in \cite{Wang} is reported on the development of  SWCNTs of 0.4 $nm$ diameter inside the nanochannels of porous zeolite $AlPO_4-5$  single crystals and the authors are also considered 0.4 $nm$ diameter NTs to
be  one-dimensional quantum systems. Given tubes were shown to exhibit unusual
novel phenomena like diamagnetism and superconductivity at low temperatures \cite{Tang}. It seems to be the consequence of quasionedimensionality of given NTs.
The additional direct confirmation of quasionedimensionality in relation to small diameter NTs are recent electron spin resonance (ESR) studies in \cite{Rao}, where   
ESR
measurements on ultra-small (0.4 nm) single walled carbon nanotubes embedded in a SAPO 5
zeolite matrix with a main point of attention to potentially occurring CESR (ESR on conductivity electrons in C-band) signals with g-value strongly deviated from free electron g-value g = 2.0023 and equaled to typical for 2D metallic tubes g-value in the range $(g = 2.05 - 2.07)$ \cite{Chauvet}. Instead, the only one paramagnetic signal was observed of symmetric shape at $g = 2.0025$ on the CNTs in zeolite cages, which, how it is argued in \cite{Yearchuck_Dovlatova} corresponds to deep level states. Consequently, it is indication of the appearance of bandgap in starting metal tubes, which is typical for 1D metals, being  to be the consequence of Peierls metal-semiconductor transition. By the way, it can mean, that 2D-classification  for 1D NTs has to be modified.

The paper is organized in the following way. In Sec.2 the results and discussion are presented.  In Sec.3 the conclusions are formulated.
\section {Results and Discussion}
In quasi-1D-case single
 ZSCNT  can be considered  to be the set of carbon backbones of trans-polyacetylene ($t$-PA) chains,  which are connected between themselves by usual bonds of honeycomb graphene lattice with cylinder symmetry curvature. It is substantial, that  the theory  of electronic properties of  $t$-PA chains is very well developed and it has very good experimental confirmation. Let us touch on given subject in more detail. Pioneering works \cite{Su_Schrieffer_Heeger_1979}, \cite{Su_Schrieffer_Heeger_1980} were opening  new era in the physics of conjugated 1D-conductors. The authors have found the most simple way to describe mathematically rather complicated system - the chain of $t$-PA,  vibronic system of which represents itself the example of Fermi liquid. It is now well known SSH-model. The most substantial suggestion in SSH-model is the suggestion, that the only dimerization coordinate $u_n$ of the $n$-th $CH$-group, 
$n = \overline{1,N}$ along chain molecular-symmetry axis $x$ is substantial for determination of main physical properties of the material. Other five degrees of freedom were not taken into consideration. Nevertheless, the model has obtained magnificent experimental confirmation. Naturally, it is reasonable to suggest, that given success is the consequence of some general principle.   Given general principle is really exists and main idea was proposed by Slater at the earliest stage of quantum physics era already in 1924. It is - "Any atom may in fact be supposed to communicate with other atoms all the time it is in stationary state, by means of virtual field of radiation, originating from oscillators having the frequencies of possible quantum transitions..." \cite{Slater}. Given idea will obtain its development, if to clarify  the origin  of virtual field of radiation. It is shown in \cite {Dovlatova}, that Coulomb field in 1D-systems or 2D-systems has the character of radiation field and it can exist without the sources, which have created given field. Consequently, Slater principle can be applied to t-PA and small NTs, for which Coulomb field can be considered to be "virtual" field, since both the materials are quasi-1D.  It produces preferential direction in atom  communication the only along chain axis (to be consequence of quasionedimensionality), and it explains qualitatively the success of SSH-model. 

It is remarkable, that SSH-model along with the physical basis of the existence of solitons, polarons, breathers, formed in $\pi$-electronic subsystem ($\pi$-solitons, $\pi$-polarons, $\pi$-breathers) contains in implicit form  also the basis for the existence of similar quasiparticles in  $\sigma$-electronic subsystem, discussed in \cite{Yearchuck_PL}, \cite{Yearchuck_Dovlatova}. 
The cause  is the same two-fold degeneration of ground state  of the whole electronic system, which has the form of Coleman-Weinberg potential with two minima at the values of dimerization coordinate $u_0$ and $ -u_0$, which leads to dimeriszation of interatomic distance. It means, that along with dimerization of $\pi$-bonds, the $\sigma$-bonds will also be dimerized in the same manner, which is the reason of the appearance of topological defects in $\sigma$-subsystem, that is $\sigma$-solitons, $\sigma$-polarons and $\sigma$-breathers. 

The shapes, for instance, of $\pi$-solitons and
$\sigma$-solitons can be given by the expression with the same mathematical form
\begin{equation}
\label{Eq1}
|\phi(n)|^2 = \frac{1}{\xi_{\pi(\sigma)}} sech^{2}[\frac{(n-n_0)a}{\xi_{\pi(\sigma)}} - v_{\pi(\sigma)} t] \cos \frac{n \pi}{2},
\end{equation} 
where $n, n_0$ are variable and fixed numbers of $CH$-unit in $CH$-chain, $a$ is $C-C$ interatomic spacing projection on chain direction, $v_{\pi(\sigma)}$ is $\pi$($\sigma$)-soliton velocity, $t$ is time, $\xi_{\pi(\sigma)}$ is $\pi$($\sigma$) coherence length. It is seen, that $\pi$-solitons and $\sigma$-solitons differs in fact the only by numerical value of coherence length. Given difference can be evaluated even without numerical calculation of the relation, which determines the shift of ground state energy of extended system by presence of localized perturbation. Actually it is sufficient  to take into account the known value of $\xi_\pi$ and relationships \cite{Lifshitz}
\begin{equation}
\label{Eq2}
\xi_{0\pi} = \frac{\hbar v_F}{\Delta_{0\pi}}, \xi_{0\sigma} = \frac{\hbar v_F}{\Delta_{0\sigma}},
\end{equation}
where $\Delta_{0\sigma}$, $\Delta_{0\pi}$ are $\sigma-$ and $\pi$-bandgap values at $T = 0 K$, $v_F$ is Fermi velocity.
Theoretical value $\xi_\pi$ in t-PA is $7a$, and it is low boundary in the range 
$7a - 11a$, obtained for $\xi_\pi$ from experiments \cite{Heeger_1988}. Taking into account the relationships (\ref{Eq2}),  using the value $\frac{\Delta_{\sigma}}{\Delta_{\pi}}\approx 8.8$, which was evaluated from t-PA band structure calculation in \cite{Grant}, and mean experimental value of coherence length $\overline{\xi_{\pi}} = 9a$ we obtain the value $\overline{\xi_{\sigma}} \approx 0.125 nm$. It means, that halfwidth of space region, occupied by $\sigma$-soliton in t-PA is $\approx 0,25 nm$, that is SSH-$\sigma$-solitons are much more localized in comparison with SSH-$\pi$-solitons. Similar conclusion takes place for SSH-$\sigma$-polarons representing itself the soliton-antisoliton pair.   SSH-$\sigma$-polarons have recently been experimentally detected in \cite{Yearchuck_PL}, where the formation of polaron lattice (PL) was
proposed. It was established, that  two components of each elementary unit, that is, of each polaron, possess by two equal in values electical own dipole  moments, proportional to spin, which was called electical spin moments (ESM), with opposite directions. It was shown, that experimental results agree well with  proposal of PL-formation, which means in fact  
 the formation of antiferroelectrically ordered lattice of quasiparticles. Given lattice consists of 2 sublattices, corresponding to soliton and antisoliton components of polaron. Corresponding chain state is optically active and it is characterized by  the set of lines in IR-spectra, which were assigned with new phenomenon - antiferroelectric spin wave resonance (AFESWR). Central mode is convential antiferroelectric  resonance (AFR)  mode, its value $ \nu^\sigma_p(C)$ in carbyne sample studied was 477 $cm^{-1}$. Let us remember that carbynes are organic quasionedimensional conductors with the simplest, consisting the only of the carbon atoms, chain structure. At the same time the presence of two electronic  $\pi_x$ and $\pi_y$-subsystems, which are "hung" on $\sigma$-subsystem means thatthe ground electronic state is similar to twodimensional Coleman-Weinberg potential with four minima at the values of dimerization coordinate $u_0$ and $ -u_0$. In other words, ground electronic state in carbynes is four-fold degenerate, which leads to a substantially more rich spectrum of possible quasiparticles, discussed in \cite{Rice}.

      It is also remarkable, that Slater principle is applicable  to  perfect quasi-1D zigzag NTs  despite the strong interaction (usual chemical bonds) between the chains.  Experimental confirmation for given conclusion follows from ESR-studies in rather perfect CZSNTs, produced by high energy ion modification  of diamond single crysatals, by which the appearance of  Peierls transition and SSH-soliton formation was established \cite{Ertchak}. Therefore, perfect quasi-1D zigzag NTs, characterized by $(m,0)$ indices, will have bandgap like to semiconductors  at any $m$, including the case $m = 3q, q \in N$, for which the 2D theory predicts the metallic properties.
Given difference is consequence of that, that all known to our knowledge theoretical works do not take into account the 2D-1D transition by decrease of NT-diameter, accompanying by Peierls transition for metal NTs, see for instance \cite{Dresselhaus}, \cite{Bose}.  

To observe  optical quantum coherent effects on NTs,  the ensemble of NTs has to be homogeneous. Moreover, to satisfy a requirement of onedimensionality and homogeneity any dispersion in axis direction, chirality, length and especially in diameter both for single NT along its axis and between different NTs in ensemble has to be absent, axial symmetry has to be also retained, that is, there are additional requirements in comparison with, for example, t-PA technology. The technology, based on HEIM, discussed in review \cite{Ertchak}, satisfy given requirements.  The CVD-technology and many similar ones seem to be not satisfying to above-listed requirements at present. It means, that experimental results and their theoretical treatment will be different in both the cases, that really takes place, even for tubes of small diameters if to take into account the comparison df ESR results in \cite{Erchak}, \cite{Ertchak}, \cite{Ertchak_Stelmakh} and in \cite{Rao}. The situation seems to be analogous to some extent to the solid state physics of the same substance in single crystal and amorphous forms.

We can now consider the task put by. In the frames of SSH-model  ZSCNT represents itself autonomous dynamical system with discrete circular symmetry consisting of finite number $n\in N$ of carbon backbones of $t$-PA chains, which are placed periodically along transverse angle coordinate. Longitudinal axes $\{x_i\}, i = \overline{1,n}$, of individual chains can be directed both along element of cylinder and  along generatrix  of any other smooth figure with axial symmetry. It is taken into account, that in the frames of SSH-model,  the adjacent chains, which represent themselves a mirror of each other, will be indistinguishable. Here we will consider the only SSH-$\sigma$-polarons to be optically active centers, which produce polaron lattice. Each $\sigma$-polaron in accordance with experiment \cite{Yearchuck_PL} can be approximated like to guantum dot in \cite{Slepyan_Yerchak} by two-level qubit. Then  the Hamiltonian, proposed in given work
 can be generalized.  The insufficient for the model local field term was omitted. (Local field term seems to be playing minor role by description of $\sigma$-polarons in comparison with 	quantum dots, since size of quantum dots is greatly exceeding the size of $\sigma$-polarons). We will use the apparatus of hypercomplex $n$-numbers. Let us remember, that hypercomplex $n$-numbers are defined to be elements of commutative ring
\begin{equation}
\label{Eq5}
Z_n = C \oplus {C} \oplus{...}\oplus {C},
\end{equation}
that is, it is direct sum of $n$ fields of complex numbers $C$, $n\in N$. It means that any hypercomplex $n$-number $z$ is $n$-dimensional quantity with the components $k_\alpha \in C$, $\alpha = \overline{0,n-1}$, that is in row matrix form $z$ is
\begin{equation}
\label{Eq6}
z = [k_0, k_1, k_2, ..., k_{n-1}],
\end{equation}
 it can be represented also in the form
\begin{equation}\label{Eq7}
z = \sum_{\alpha = 0}^{n-1}k_\alpha\pi_\alpha,
\end{equation}
where $\pi_\alpha$ are basis elements of $Z_n$ (and simultaneously  basis elements of the linear space of $n$-dimensional lines and  $n$-dimensional row matrix). They are
\begin{equation}
\label{Eq8}
\begin{split}
&\pi_0 = [1,0, ...,0,0],  \pi_1 =[0,1, ...,0,0],\\
&..., \pi_{n-1} = [0,0, ...,0,1].
\end{split}
\end{equation}
Basis elements $\pi_\alpha$ possess by projection properties
\begin{equation}
\label{Eq9}
\pi_\alpha\pi_\alpha = \pi_\alpha\delta_{\alpha\beta}, \sum_{\alpha = 0}^{n-1}\pi_\alpha = 1, z \pi_\alpha = k_\alpha\pi_\alpha
\end{equation}
In other words, the set of $k_\alpha \in C, \alpha = \overline{0, n-1}$ is the set of eigenvalues of hypercomplex $n$-number $z \in Z_n$, the set  of $\{\pi_\alpha\}$, $\alpha = \overline{0, n-1}$ is eigenbasis of $Z_n$-algebra.
 
Then the QED-Hamiltonian, considered to be hypercomplex operator $n$-number, for $\sigma$-polaron system of interacting with EM-field ZSCNTs, consisting of $n$ backbones of $t$-PA chains, which are connected between themselves in that way, in order to produce rolled up graphene sheet in matrix representation is \begin{equation} 
\label{Eq10}
[\hat{\mathcal{H}}] = [\hat{\mathcal{H}}_{\sigma}] + [\hat{\mathcal{H}}_F] + [\hat{\mathcal{H}}_{\sigma F}] + [\hat{\mathcal{H}}_{\sigma \sigma}].
\end{equation}
The rotating wave approximation and  the single-mode approximation of EM-field are used. All the components in (\ref{Eq10}) are considered to be hypercomplex operator $n$-numbers and they are the following. $[\hat{\mathcal{H}}_{\sigma}]$ represents the operator of $\sigma$-polaron subsystem energy in the absence of interaction between $\sigma$-polarons  themselves and with EM-field. It is
\begin{equation} 
\label{Eq11}
[\hat{\mathcal{H}}_{\sigma}] = (\hbar \omega _0/2) \sum_{j = 0}^{n-1}\sum_m {\hat {\sigma}^z_{mj}}[e_1]^j,
\end{equation}
  where $\hat {\sigma}^z_{mj} = \left|a_{mj}\right\rangle  \left\langle a_{mj} \right|-\left|b_{mj}\right\rangle  \left\langle b_{mj} \right|$ is $z$-transition operator between the ground and excited states of $m$-th $\sigma$-polaron in $j$-th chain. The second term 
\begin{equation} 
\label{Eq12}
[\hat {\mathcal{H}}_F] = \hbar \omega \sum_{j = 0}^{n-1}\hat {a}^+\hat {a}[e_1]^j 
\end{equation}
is the Hamiltonian of the free EM-field, which is represented in the form of 
hypercomplex operator $n$-number.
The component of the Hamiltonian (\ref{Eq10})
\begin{equation}
\label{Eq13}
[\hat {\mathcal{H}}_{\sigma F}] =\hbar g \sum_{j = 0}^{n-1}\sum\limits_m {(\hat {\sigma }_{mj}^+\hat {a}e^{ikma} + \hat {\sigma }_{mj}^-\hat {a}^+e^{-ikma})}[e_1]^j 
\end{equation}
describes the interaction of $\sigma$-polaron sybsystem with EM-field, where $g$  is the interaction constant.
The Hamiltonian
\begin{equation}
\begin{split}
\label{Eq14}
&[\hat{\mathcal{H}}_{\sigma\sigma}] =
-\hbar \sum_{l = 0}^{n-1}\sum_{j = 0}^{n-1}\xi^{(1)}_{|l-j|}[e_1]^l\sum\limits_m \left|a_{mj} \right\rangle \left\langle a_{m+1,j}\right|[e_1]^j \\
&-\hbar\sum_{l = 0}^{n-1}\sum_{j = 0}^{n-1}\xi^{(1)}_{|l-j|}[e_1]^l\sum\limits_m \left|a_{mj} \right\rangle \left\langle a_{m-1,j}  \right| [e_1]^j\\
&-\hbar\sum_{l=0}^{n-1}\sum_{j = 0}^{n-1}\xi^{(2)}_{|l-j|}[e_1]^l\sum\limits_m  \left| b_{mj} \right\rangle \left\langle b_{m+1,j}\right|[e_1]^j\\ &-\hbar\sum_{l=0}^{n-1} \sum_{j = 0}^{n-1}\xi^{(2)}_{|l-j|}[e_1]^l\sum\limits_m\left| b_{mj} \right\rangle \left\langle b_{p-1,j} \right|[e_1]^j,
\end{split}
\end{equation}
where  $\hbar\xi^{(1,2)}_{|l-j|}$ are the energies, characterizing intrachain $(l = j)$ and interchain $(l \neq j)$ polaron-polaron interaction for the excited ($\xi^{(1)}$) and ground ($\xi^{(2)}$) states of $j$-th chain, $\left| b_{mj} \right\rangle, \left|a_{mj}\right\rangle$ are ground and excited states correspondingly of $m$-th $\sigma$-polaron of $j$-th chain. In (\ref{Eq11}) to (\ref{Eq14}) $[e_1]^j$ is j-th power of the circulant matrix $[e_1]$, which is
\begin{equation}
\label{Eq16}
[e_1]=\left[\begin{array} {*{20}c} 0&1&0& ...&0  \\ 0&0&1& ...&0 \\ &...& \\ 0&0& ... &0&1\\1&0&...&0&0 \end{array}\right].
\end{equation}
Hamiltonian in the form like to (\ref{Eq14}) at $n = 1$ is usually used for description of tunneling 
between the states with equal energies, in particular, for tunneling between quantum dot states \cite{Slepyan_Yerchak}. Hamiltonian (\ref{Eq14}) at any $n$ describes actually the connection between pairs of the states, which satisfy the following condition - the first state  in any pair results from the second state (and vice versa)  by time reversal. It is known, that for given states Cooper effect takes place. Therefore,  the application of Hamiltonian like to (\ref{Eq14}) is possible for any pair of time reversal symmetric states with equal energy.  
By the way, if one
omits last  term in Hamiltonian (\ref{Eq10}), it goes into $n$-chain generalization of
well-known Tavis-Cummings Hamiltonian.
\begin{figure}
\includegraphics[width=0.5\textwidth]{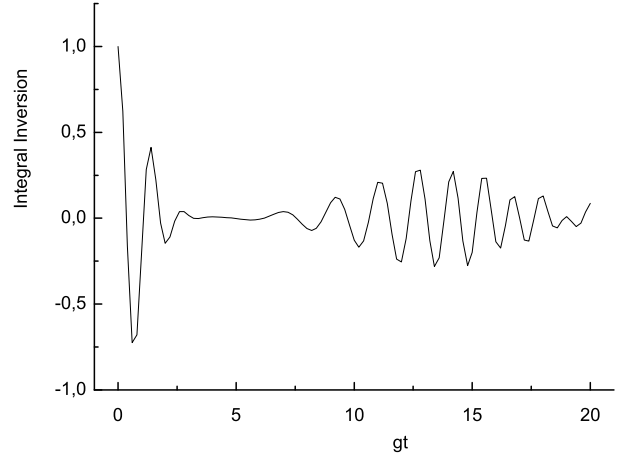}
\caption[Temporal dependence of the integral inversion  in the QD chain for a coherent initial state of light]
{\label{Figure1} Temporal dependence of the integral inversion  in the QD chain for a coherent initial state of light ($\left\langle n\right\rangle=4$).}
\end{figure}
\begin{figure}
\includegraphics[width=0.5\textwidth]{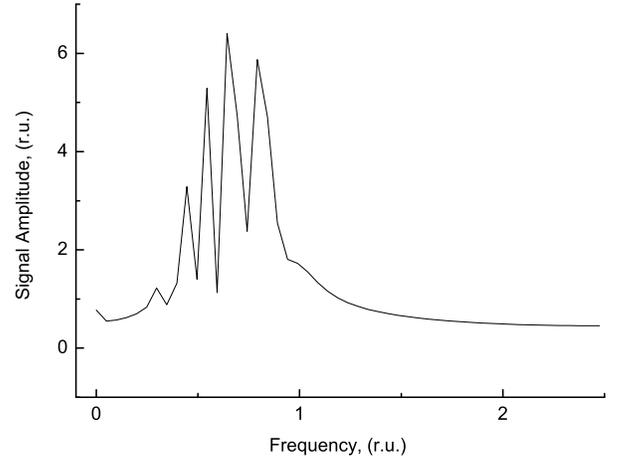}
\caption[Spectral dependence of the signal amplitude, corresponding to  temporal dependence, presented in Fig.1]
{\label{Figure2} Spectral dependence of the signal amplitude, corresponding to  temporal dependence, presented in Fig.1}
\end{figure}

 We can represent the state vector of the "NT+EM-field" system in terms of the eigenstates of isolated polaron and photon number states in the folowing matrix form
\begin{equation}
\begin{split}
\label{Eq17}
&[\left| {\Psi (t)} \right\rangle] = \\ 
&\sum_{j = 0}^{n-1}\{\sum\limits_l \sum\limits_{m} \left(A^j_{m,l}(t) \left|a_{mj},l \right\rangle + B^j_{m,l}(t) \left| b_{mj},l \right\rangle\right) \}[e_1]^j.
\end{split}
\end{equation}
Here, $\left| b_{mj},l \right\rangle = \left| b_{mj} \right\rangle\otimes\left|l \right\rangle$, $\left| a_{mj},l \right\rangle = \left| a_{mj} \right\rangle\otimes\left|l \right\rangle$, where  $\left|l \right\rangle$ is the EM-field  Fock state with $l$  photons, $A^j_{m,l}(t)$, $B^j_{m,l}(t)$ are the unknown probability amplitudes. Let us pay the attention,  that matrix function  $ [\left| {\Psi (t)} \right\rangle]$ is one state and the representation in the form of sum over j is like to representation of complex-valued wave function $ \Psi (t)$  in the form $ \Psi (t) = \Psi_1 (t) + i \Psi_2 (t)$  of two real-valued functions.
Nonstationary Schr\"odinger equation for hypercomplex matrix function $[\left| {\Psi (t)} \right\rangle]$ in the interaction representation is  
\begin{equation}
\begin{split}
\label{Eq17a}
 i\hbar \frac{\partial}{\partial{t}}[\left| {\Psi (t)} \right\rangle] = [\hat{V}(t)][\left| {\Psi (t)} \right\rangle],
\end{split}
\end{equation}
where matrix Hamiltonian of interaction $[\hat{V}(t)]$ is
\begin{equation}
\begin{split}
\label{Eq17b}
&[\hat{V}(t)] = \exp(\frac{i}{\hbar}[\hat{\mathcal{H}}_F] t)([\hat{\mathcal{H}}_{\sigma}] + [\hat{\mathcal{H}}_{\sigma F}] + \\
&[\hat{\mathcal{H}}_{\sigma \sigma}])\exp(-\frac{i}{\hbar}[\hat{\mathcal{H}}_F] t)
\end{split}
\end{equation}
Consequently, for matrices of probability amplitudes 
\begin{equation}
\begin{split}
\label{Eq17bc}
&[A_{m,l}(t)] = \sum_{j = 0}^{n-1}A^j_{m,l}(t)[e_1]^j,\\
&[B_{m,l}(t)] = \sum_{j = 0}^{n-1}B^j_{m,l}(t)[e_1]^j
\end{split}
\end{equation}
 we obtain the following system of matrix equations
\begin{equation}
\begin{split}
\label{Eq17c}
&\frac{\partial}{\partial{t}}[A_{m,l}(t)] = -\frac{i}{2} \omega_0 [A_{m,l}(t)] + i [\xi^{(1)}] ([A_{m-1,l}(t)] \\ 
&+ [A_{m+1,l}(t)]) - ig \sqrt{l + 1}[B_{m,l+1}(t)]e^{i(kma - \omega t)},
\end{split}
\end{equation}
\begin{equation}
\begin{split}
\label{Eq17d}
&\frac{\partial}{\partial{t}}[B_{m,l+1}(t)] = -\frac{i}{2} \omega_0 [B_{m,l+1}(t)] + i [\xi^{(2)}] ([B_{m-1,l+1}(t)] \\
&+ [B_{m+1,l+1}(t)]) - ig \sqrt{l + 1}[A_{m,l}(t)]e^{-i(kma - \omega t)}.
\end{split}
\end{equation}
Here
$[\xi_1]$, $[\xi_2]$ are $[n \times n]$ matrices of coefficients, determined by (\ref{Eq14}), that is
\begin{equation}
\begin{split}
\label{Eq24}
[\xi^j_1] = \sum_{l = 0}^{n-1}\xi^{(1)}_{|l-j|}[e_1]^l, [\xi^j_2] = \sum_{l = 0}^{n-1}\xi^{(2)}_{|l-j|}[e_1]^l,
\end{split}
\end{equation}
where one takes into account, that in view of axial symmetry $[\xi^j_{1,2}]$ do not depend on $j$. 
Consequently, we have $[\xi^j_1] = [\xi_1]$, $[\xi^j_2] = [\xi_2]$.

We will take into consideration the relaxation processes by means of standard  phenomenological way like to  \cite{Slepyan_Yerchak}, that is  by substituting instead 
real value $\omega_0$ the complex values  $\omega_0 - i\lambda$ and $\omega_0 +   i\lambda$ in the equations (\ref{Eq17c}) and (\ref{Eq17d}) correspondingly. It is suggested, that relaxation time $\tau$ is independent on the chain number and it is determined by the value, which is reciprocal to $\lambda$, that is  $\tau = \frac{2\pi}{\lambda}$.

Let us define block matrix
\begin{equation}
\begin{split}
\label{Eq17e}
[\Psi_{m,l}(t)] = \left[\begin{array} {*{20}c}&[A_{m,l}(t)]\\
&      \\
&[B_{m,l+1}(t)]\end{array}\right],  
\end{split}
\end{equation} 
consisting of two $[n \times n]$ matrices of probability amplitudes
\begin{equation}
\begin{split}
\label{Eq21}
&[A_{m,l}(t)] = \sum_{j = 0}^{n-1}A^j_{m,l}(t)[e_1]^j, \\
&[B_{m,l+1}(t)] = \sum_{j = 0}^{n-1}B^j_{m,l+1}(t)[e_1]^j,
\end{split}
\end{equation}
which are determined by relationship (\ref{Eq17}).
Then the set of difference-differencial matrix equations (\ref{Eq17c}) and (\ref{Eq17d}) can be rewritten in compact form
\begin{equation}
\begin{split}
\label{Eq17f}
&\frac{\partial}{\partial{t}}[\Psi_{m,l}(t)] = \{-\frac{i}{2} \omega_0 [\sigma_z] - \frac{\lambda}{2}[E] - \\  
&i g \sqrt{l + 1}[\sigma_z] \exp({i[\sigma_z] (\omega t - kma)})\}\  
[\Psi_{m,l}(t)] + \\  
&i [\xi] ([\Psi_{m-1,l}(t)] + [\Psi_{m+1,l}(t)]),
\end{split}
\end{equation}
where $[\sigma_z]$ is Pauli $z$-matrix, $[E]$ is $[2 \times 2]$ unit matrix, $[\xi]$ is block matrix
\begin{equation}
\begin{split}
\label{Eq17g}
[\xi] = \frac{1}{2} [(\xi^{(1)} + \xi^{(2)})\otimes[E] + (\xi^{(1)} - \xi^{(2)})\otimes[\sigma_z]].
\end{split}
\end{equation}

 The solution of hypercomplex equation (\ref{Eq17f}) was obtained   in continuum limit. For the state vector $[\left| {\Psi (t)} \right\rangle]$   in continuum limit we have
\begin{equation}
\begin{split}
\label{Eq18}
&[\Phi^l(x,t)] = \\
&\int\limits_{-\infty }^\infty{[\overline{\Phi}^l(h,0)]\exp\{i t([\theta^l(h)] - g \sqrt{l+1}[\chi])\}e^{ihx}}dh,
\end{split}
\end{equation}
where $x$ is  hypercomplex axis $x = [x, x, ..., x]$, $[\Phi^l(x,t)]$ is 
\begin{equation}
\begin{split}
\label{Eq19}
[\Phi^l(x,t)] = exp{\frac{i(\omega_0 t - kx)[\sigma_z]}{2}}\exp{\frac{\lambda t}{2}}[\Psi^l(x,t)], 
\end{split}
\end{equation}
 In its turn $[\Psi^l(x,t)]$ is continuous limit of functional block matrix of discrete variable $m$, which is given by (\ref{Eq17e})
Further, matrix $[\theta(h)]$ in (\ref{Eq18}) is 
\begin{equation}
\begin{split}
\label{Eq22}
&[\theta(h)] = \frac{1}{2}\{([\theta_1(h)] + [\theta_2(h)]) \otimes [E_2]\} \\
&+ \frac{1}{2}\{([\theta_1(h)] - [\theta_2(h)]) \otimes [\sigma_z]\}, 
\end{split}
\end{equation}
where $[E_2]$ is unit $[2 \times 2]$-matrix, $[\theta_1(h)]$ and $[\theta_2(h)]$ are 
\begin{equation}
\begin{split}
\label{Eq23}
&[\theta_1(h)] = [\xi_1]\{2 - a^2(h + \frac{k}{2})^2\}, \\
&[\theta_2(h)] = [\xi_2]\{2 - a^2(h - \frac{k}{2})^2\}
\end{split}
\end{equation}
Here  $[\xi_1]$, $[\xi_2]$ are $[n \times n]$ matrices of coefficients, defined by  (\ref{Eq24}). Matrix $[\chi]$ in (\ref{Eq18}) is
\begin{equation}
\begin{split}
\label{Eq25}
[\chi] = [E_n] \otimes [\sigma_x]\exp ({-i [\sigma_z] (\omega - \omega_0) t}).
\end{split}
\end{equation}
Matrix elements of $[\Phi^l(x,t)]$ are
\begin{equation}
\begin{split}
\label{Eq26}
&\Phi^l_{qp}(x,t) = \int\limits_{-\infty }^{\infty}\Theta^l_{q}(h,0) \exp{ \frac{-2\pi qpi}{n}} \exp{ihx}\times \\
&\exp{\{i\sum_{j = 0}^{n-1}\exp{\frac{2\pi qji}{n}(\vartheta_j(h) - g\sqrt{l-1}\kappa_j(h))}\}}dh,
\end{split}
\end{equation}
where $\Theta^l_{q}(h,0)$,$\vartheta_j(h)$, $\kappa_j(h)$ are determined by eigenvalues $\textbf{k}_\alpha \in C, \alpha = \overline{0, n-1}$ of $\Phi^l(h,0)$, $\theta(h)$ and $\chi(h)$, which are considered to be $n$-numbers. They are
\begin{equation}
\label{Eq27}
\Theta^l_{q}(h,0) = \frac{1}{n}\textbf{k}_q (\Phi^l(h,0)) = \frac{1}{n}\sum_{j = 0}^{n-1}\Phi_j^l(h,0)\exp{\frac{2\pi q j i}{n}}
\end{equation}
\begin{equation}
\label{Eq28}
\vartheta_j(h) = \frac{1}{n}\textbf{k}_j(\theta(h)) = \frac{1}{n}\sum_{r = 0}^{n-1}\theta_r(h)\exp{\frac{2\pi  jr i}{n}},
\end{equation}
\begin{equation}
\label{Eq29}
\kappa_j(h) = \frac{1}{n}\textbf{k}_j(\chi(h)) = \frac{1}{n}\sum_{r = 0}^{n-1}\chi_j(h)\exp{\frac{2\pi j r i}{n}}.
\end{equation}
Then the hypercomplex solution  can be represented in the form of sum of $n$ solutions for $n$ chains, that is, hypercomplex $n$-number $\Phi^l(x,t) $ is
\begin{equation}
\label{Eq30}
\Phi^l(x,t) = \sum_{q = 0}^{n-1}\tilde{\Phi}^l_q(x,t),
\end{equation}
where the solution for $q$-th chain $\tilde{\Phi}^l_q(x,t)$ is
\begin{equation}
\label{Eq31}
\tilde{\Phi}^l_q(x,t) = \sum_{p = 0}^{n-1}\Phi^l_{qp}(x,t)[e_1]^p,
\end{equation}
in which the  matrix elements $\Phi^l_{qp}(x,t)$ are  determined by (\ref{Eq26}).
The relationship (\ref{Eq31}) by taking into account  (\ref{Eq18}) - (\ref{Eq30}) determines Rabi-wave packet, which propagates along individual chain of zigzag NT.  It is also clear, that subsequent analysis of Rabi-wave packet dynamics for individual NT-component will be the same (by rescaling of parameters), that in \cite{Slepyan_Yerchak}. The parameters can in principle be obtained by detailed comparison of results obtained with experiments on optical absorption reflection or Raman scattering. It is the subject for futher work. For qualitative analysis we can use arbitrary parameters of the task, in particular for the convenience the parameters, used in  \cite{Slepyan_Yerchak} for quantum dot (QD) chain. They are:  a coherent initial state of light with $\left\langle n\right\rangle=4$, the initial state of QD-chain is a single Gaussian wavepacket, that is
$A_n(x,0)=c(n)\exp(-x^2/2\sigma^2)/\sqrt[4]{\pi\sigma^2}$, $B_{n+1}(x,0)=0$, parameters
$\xi_1=10g$, $\xi_2=7g$, deviation from resonance is choosed to be equal $\omega_0 - \omega$ = $\Delta = 2(\xi_1-\xi_2)+\xi_2a^2k^2$,
$ka=0.5$, $\sigma = 20 a$,  $\lambda = 0.05g$. Then we will obtain the same temporal dependence of the integral inversion, Figure 1, coinciding with  Figure 4 in \cite{Slepyan_Yerchak}. We wish to demomstrate only, that if, in particular, temporal dependence of the integral inversion for QD-chain,  or for quasi-1D ZSCNT corresponds  to Figure 1, then optical spectra  will  have the form, presented in Figure 2.  hat an originally Gaussian packet temporally oscillates, at that oscillations collapse to zero quickly, but revive with time increasing in, and what is characteristic, in another area of space \cite{Slepyan_Yerchak}. 

In other words, temporal dependence, Figure 1, of the integral inversion gives in implicit form  the way for  comparison of theoretical results with any stationary optical experiments in QD chain or like them, including quasi-1D ZSCNTs, with aforesaid initial state. 
Really, it is sufficient to make a Fourier transform of  given temporal dependence, Figure 2. It will be proportional to signal amplitudes of  infrared (IR) absorbance, IR-transmittance, IR-reflectance or Raman scattering, since they are  determined by population difference. It means, that dynamical nonstationary properties of optical systems can become apparent by conventional stationary registration of the spectra. It is very  similar to the well known in stationary ESR-spectroscopy situation, where the signals of the centers with long relaxation times can be registered in nonstationary regime, which can lead to zeroth absorption (or  appearance of the signals in inverse phase like to  maser effects), that was practically used in very many ESR-studies to unravel complicated overlapped spectra (however without theoretical explanation of given effect). At the same time there is difference of principle between classically considered nonstationary regime with Rabi oscillations and between QED-consideration.  Classical consideration indicates, that  operating point by the registration of the centers with long relaxation times can move  along Rabi oscillation curve, leading to radical changes in amplitude of the signals, including the appearance of irradiation instead absorption, naturally in given process the energy of EM-field source is partly  came back from spin systems in the same quality (in EM-field form) instead its transformation in phonon energy. Given process can be accompanying by enhanced noice level, however without appearance of any additional lines. QED-consideration shows,  that the process  of Rabi waves' formation  is determined essentially by interaction  of optical centers with photon subsystem, at that stationary regime is achieved fastly, giving usual stationary optical signals, however after a time the oscillations emerge again, and given revival part leads to additional lines in high energy part of spectral dependence. Therefore, the  absorption (scattering) process by presence of Rabi waves is always nonstationary process and it can be realized even at relatively short relaxation times of optical centers, constant $g$ of the interaction of optical centers with EM-field is large.

Therefore, the NT-QED-model predicts the appearance of additional lines, for which Rabi waves are responsible, in stationary optical spectra of quasi-1D ZSCNTs. It is understandable, that   Rabi waves can be  registered  in any 1D-systems by stationary optical measurements, if the electron-photon coupling is rather strong. 
The  experimental confirmation of  results presented is given in \cite{Yearchuck_Dovlatova}. 
\section{Conclusions}
Thus, in fact, we have obtained for the first time in QED the analytical solution of the task of the interaction of quantized EM-field with multichain qubit quasi-1D axially symmetric system by taking into account both the intrachain and interchain qubit coupling.
The model considered continues the QED-model series,  in which the next model generalizes previous model and which at the same time can be solved analytically, that is,  it ranks on a par with 
Jaynes-Cummings model  for the one qubit, Tavis-Cummings model, which is JCM-generalization for multiqubit systems and  the model described in \cite{Slepyan_Yerchak}, generalizing  Tavis-Cummings model  by taking into account the 1D-coupling between qubits. 

The appearance of additional lines in optical spectra of aforesaid quasi-1D  systems  strongly interacting with EM-field is predicted. They are result of nonstationary registration conditions, to be consequence of Rabi wave packet formation and correspond to Fourier transformation of revival part of Rabi wave packets in temporal dependence of the integral inversion. The formation of Rabi wave packets is in its turn the consequence of quantum nature of EM-field.

 It is argued the applicability of Su-Schrieffer-Heeger-model of conjugated organic conductors for perfect  quasi-1D carbon zigzag shaped nanotubes.  The applicability of Su-Schrieffer-Heeger-model is consequence of 2D-1D transition  in  quasi-1D carbon zigzag shaped nanotubes with diameter decrease, resulting in cardinal change of all their physical properties.

\end{document}